\documentclass[a4paper,11pt]{article}

\usepackage{pos}

\title{Galactic Core-Collapse Supernovae at IceCube: ``Fire Drill'' Data Challenges and follow-up}

\ShortTitle{IceCube Supernova Fire Drills and Follow-up}

\author{The IceCube Collaboration \\{\normalsize \normalfont(a complete list of authors can be found at the end of the proceedings)}\\}

\emailAdd{sgriswold@icecube.wisc.edu}
\emailAdd{segev.benzvi@icecube.wisc.edu}

\abstract{

The next Galactic core-collapse supernova (CCSN) presents a once-in-a-lifetime opportunity to make astrophysical measurements using neutrinos, gravitational waves, and electromagnetic radiation. CCSNe local to the Milky Way are extremely rare, so it is paramount that detectors are prepared to observe the signal when it arrives. The IceCube Neutrino Observatory, a gigaton water Cherenkov detector below the South Pole, is sensitive to the burst of neutrinos released by a Galactic CCSN at a level $>$10$\sigma$. This burst of neutrinos precedes optical emission by hours to days, enabling neutrinos to serve as an early warning for follow-up observation. IceCube's detection capabilities make it a cornerstone of the global network of neutrino detectors monitoring for Galactic CCSNe, the SuperNova Early Warning System (SNEWS 2.0). In this contribution, we describe IceCube's sensitivity to Galactic CCSNe and strategies for operational readiness, including ``fire drill'' data challenges. We also discuss coordination with SNEWS 2.0.

\vspace{4mm}
{\bfseries Corresponding authors:}
Spencer Griswold$^{1*}$, Segev BenZvi$^{1}$\\
{$^{1}$ \itshape Dept. of Physics and Astronomy, University of Rochester\\
  206 Bausch \& Lomb Hall P.O. Box 270171, Rochester, NY 14627, USA}\\[4mm]
$^*$ Presenter

\ConferenceLogo{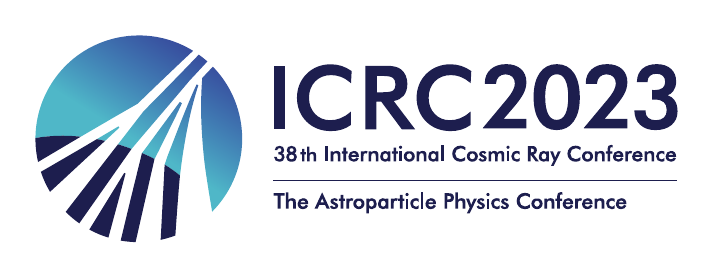}

\FullConference{The 38th International Cosmic Ray Conference (ICRC2023)\\ 26 July -- 3 August, 2023\\ Nagoya, Japan}
}

\begin{document}

\maketitle

\section{Introduction}
\begin{figure}[b]
    \centering
    \includegraphics[width=0.9\textwidth]{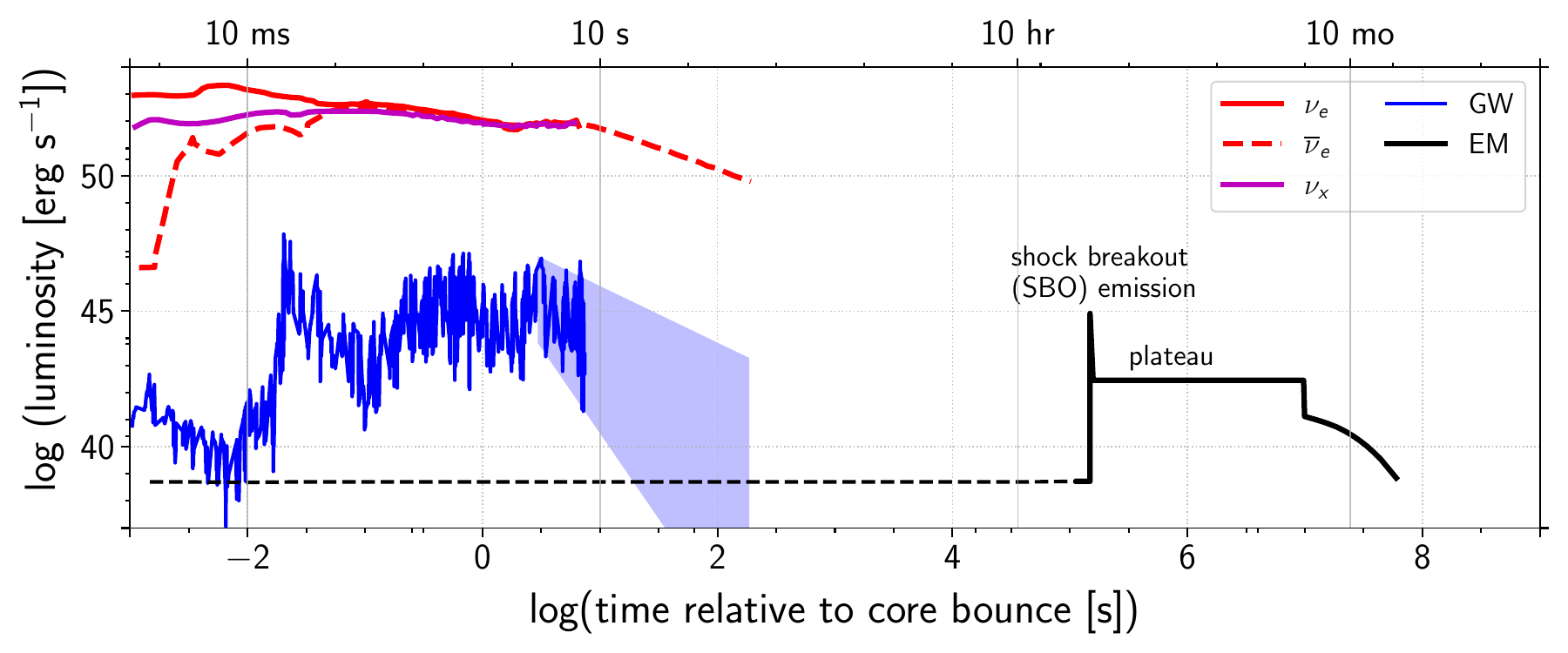}
    \caption{Luminosity of neutrinos, photons (EM), and gravitational waves (GW) originating from a 17~$M_\odot$ CCSN progenitor, adapted from \cite{Nakamura:2016kkl}.}
    \label{fig:mm_signals}
\end{figure}

A Galactic core-collapse supernova (CCSN) will produce a high luminosity burst of all flavors of neutrinos. As neutrinos are neutral and are not extinguished by interstellar dust, they serve as valuable astrophysical messengers. In particular, neutrinos produced by a CCSN may be used to probe the core structure and equation of state of the exploding star. Observing neutrinos from a CCSN would also provide insight into fundamental neutrino physics and potentially physics beyond the Standard Model \cite{johns:2021snowmass}. CCSN neutrinos will also support optical follow-ups and complimentary measurements using other astrophysical messengers. The burst of neutrinos produced in a CCSN is expected to precede optical emission by hours to days (see Fig. \ref{fig:mm_signals}) and thus can provide an early warning for optical observations, which is critical to enable observation of features such as the optical breakout burst. Gravitational waves are also expected to arrive in tandem with the neutrino burst, enabling measurement of the absolute mass of the neutrino \cite{johns:2021snowmass}. The next Galactic CCSN event presents a once-in-a-lifetime opportunity to make a multi-messenger astrophysical measurement. These events are exceedingly rare and occur in the Milky Way once every 60 years \cite{Rozwadowska:2020nab}. It is thus necessary to prepare for this event to ensure that online software, detector hardware, and experimental operators are ready to make this measurement when the opportunity arises.

IceCube, the world's largest neutrino detector, instruments a cubic kilometer of ice at the geographic South Pole using a lattice of 5,160 digital optical modules (DOMs) deployed in glacial ice between 1.5 km and 2.5 km below the surface \cite{Aartsen:2016nxy}. The DOMs are arranged on 86 strings, each with 60 DOMs, spaced 125~m apart in a hexagonal grid. On each string, DOMs are spaced out 17~m vertically. Each DOM is equipped with a 10-inch-diameter Hamamatsu photomultiplier tube (PMT) to capture Cherenkov photons produced by neutrino interactions in the ice. In the center of IceCube, a group of 8 strings is equipped with DOMs with 35\% higher quantum efficiency \cite{Collaboration:2011ym}. This sub-array, DeepCore, is arranged with an average inter-string distance of 72 m and an inter-DOM spacing ranging from 7 m to 10 m. The dense arrangement of DeepCore DOMs improves IceCube’s sensitivity to low-energy neutrinos. IceCube DOMs observe a background rate in excess of 500 Hz originating from dark current in the PMTs, afterpulses in the PMTs occurring several $\mu$s after absorbing a high-energy photon, and radioactive decays in the DOM's glass housing. Both background and signal “hits”, defined as the recording of a photon by a PMT, are tagged by IceCube’s primary physics data acquisition system (pDAQ) based on their spatial and temporal clustering in the detector. A simple multiplicity trigger is applied to hits, denoted SMTn, where n is the number of hits that must be observed within a time window tuned for each sub-array of the detector. For example, IceCube uses an SMT8 trigger condition with a coincidence window of  $\pm5$ $\mu$s. This is optimized for events with energy >100 GeV and has a trigger rate of $\sim$3 KHz, mainly due to atmospheric muons \cite{Aartsen:2016nxy}.When multiple triggers form, they are combined into a global trigger, which prompts data storage, cleaning, and event reconstruction.

A CCSN neutrino burst at the galactic center is expected to produce an all-flavor neutrino flux of $10^{16}$ m$^{-2}$  with energies $\mathcal{O}$(10 MeV) over a period of 10~s \cite{Abbasi:2011ss}. Individual MeV neutrinos will interact in the ice and the daughter leptons, such as the positron produced in inverse beta decay ($\bar{\nu}_e+p\to n+e^+$), will travel an average distance of 5 cm in the ice and produce several hundred Cherenkov photons. Due to the sparseness of the IceCube array, an average of one photon would be detected from each interaction \cite{Abbasi:2011ss}. This is insufficient to reconstruct individual neutrinos and would not be detectable over the large per-DOM background rate. However, the neutrino burst would affect all of the detector's instrumented volume, forming a detectable collective increase in the hit rates across all DOMs. IceCube is sensitive to the neutrino signal from a galactic CCSN at a significance level $\gg10\sigma$ \cite{Abbasi:2011ss}.

\begin{figure}
    \centering
    \includegraphics[width=\textwidth]{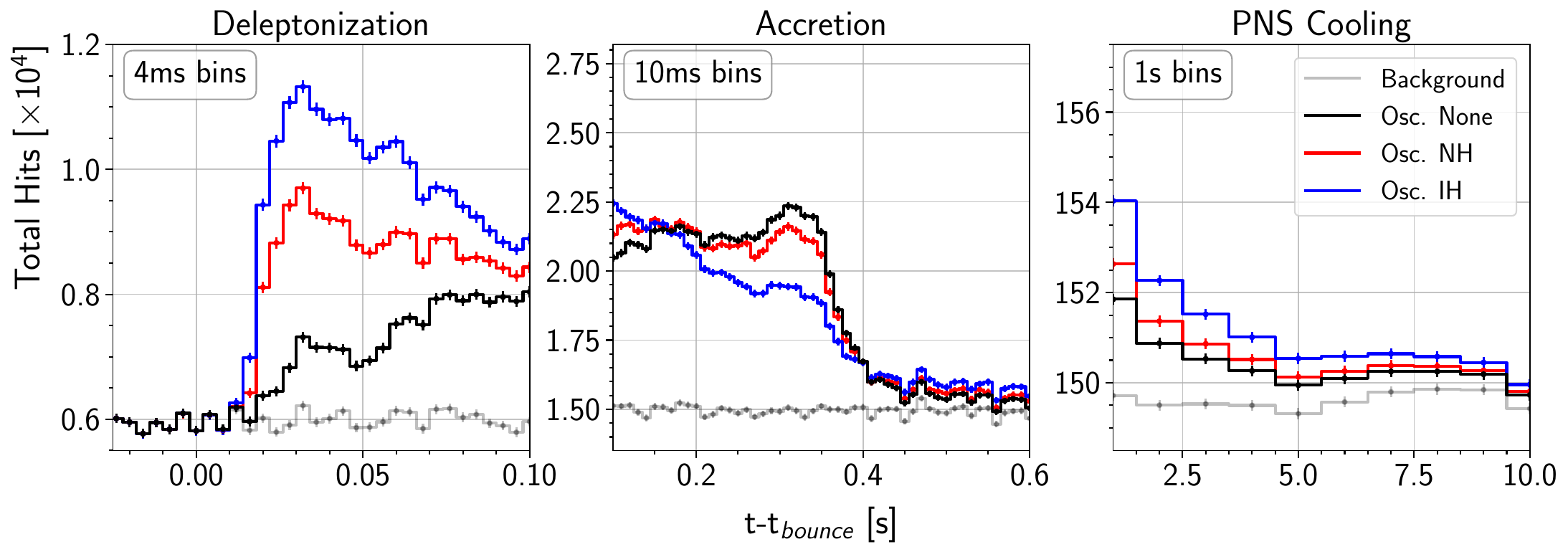}
    \caption{Simulated hits in IceCube due to a 13 M$_\odot$ progenitor 10 kpc from Earth. Model from \cite{Nakazato:2012qf}, simulated in ASTERIA \cite{ASTERIA:2019} and SNEWPY \cite{baxter2021snewpy}. Different bin sizes are chosen to illustrate time features in different phases of a CCSN \cite{Janka:2017}: (left) the deleptonization peak, (center) a $\sim0.5$~s plateau as matter accretes onto the core forming a proto-neutron star (PNS), and (right) $\sim10$~s of exponential decay as the PNS cools.}
    \label{fig:hits_phases}
\end{figure}

IceCube is particularly well-suited to monitor the Milky Way for CCSNe. Since 2015, IceCube's supernova data acquisition system (SNDAQ) has operated with $>$99\% trigger-capable uptime and is supported by the HitSpool data buffering system \cite{Heereman:2015mbs}. SNDAQ issues alerts on supernova triggers in real time, as described in Sec. \ref{sec:sn_detection}, and issues requests to HitSpool to buffer the DOM waveforms in a 90~s window surrounding the trigger time \cite{Heereman:2015mbs}. As of November 2021, approximately 13 days' worth of data is buffered by the HitSpool system, and is available for offline analysis upon request. While SNDAQ triggers using a stream of 1.6~ms bins, the waveforms provided by HitSpool enable measuring the neutrino lightcurve with ns precision. An example of the expected lightcurve observed by IceCube in response to a 13 M$_\odot$  progenitor by Nakazato et al. is provided in Fig. \ref{fig:hits_phases} \cite{Nakazato:2012qf}. 

IceCube is a key component of the SuperNova Early Warning System (SNEWS 2.0) \cite{Kharusi:2020ovw}, a network of neutrino detectors designed to give advanced notice of imminent optical emission from a nearby CCSN. By identifying coincidences between the arrival times of neutrino bursts in detectors around the world, SNEWS is intended to facilitate robust optical follow-ups of CCSNe. IceCube has coordinated its supernova alerts with SNEWS since 2009. Currently, IceCube issues alerts to SNEWS at a rate of about once per month. Since 2018, IceCube has transmitted a second diagnostic data stream of low-significance alerts to SNEWS at the rate of about 5 triggers per day. The diagnostic channel tests the latency and health of the connection between IceCube and SNEWS, and is used to test the SNEWS multi-detector coincidence software. Additionally, IceCube actively participates in the development and testing of SNEWS software including \texttt{SNEWPY}, a unified CCSN model interface \cite{baxter2021snewpy} and \texttt{SNEWS\_Publishing\_Tools} (\texttt{snews\_pt}), a CCSN alert management system \cite{snews_pt}.

In this contribution, we discuss the detection of CCSNe neutrinos at the IceCube Neutrino Observatory; the method used to test the formation of supernova triggers; and tests of IceCube's operational readiness to trigger on and respond to CCSN neutrino bursts. We also discuss future plans for coordinating data challenges with other neutrino detectors and optical telescopes through SNEWS 2.0.

\section{Supernova detection at IceCube}
\label{sec:sn_detection}
\begin{figure}[b]
    \centering
    \includegraphics[height=6cm]{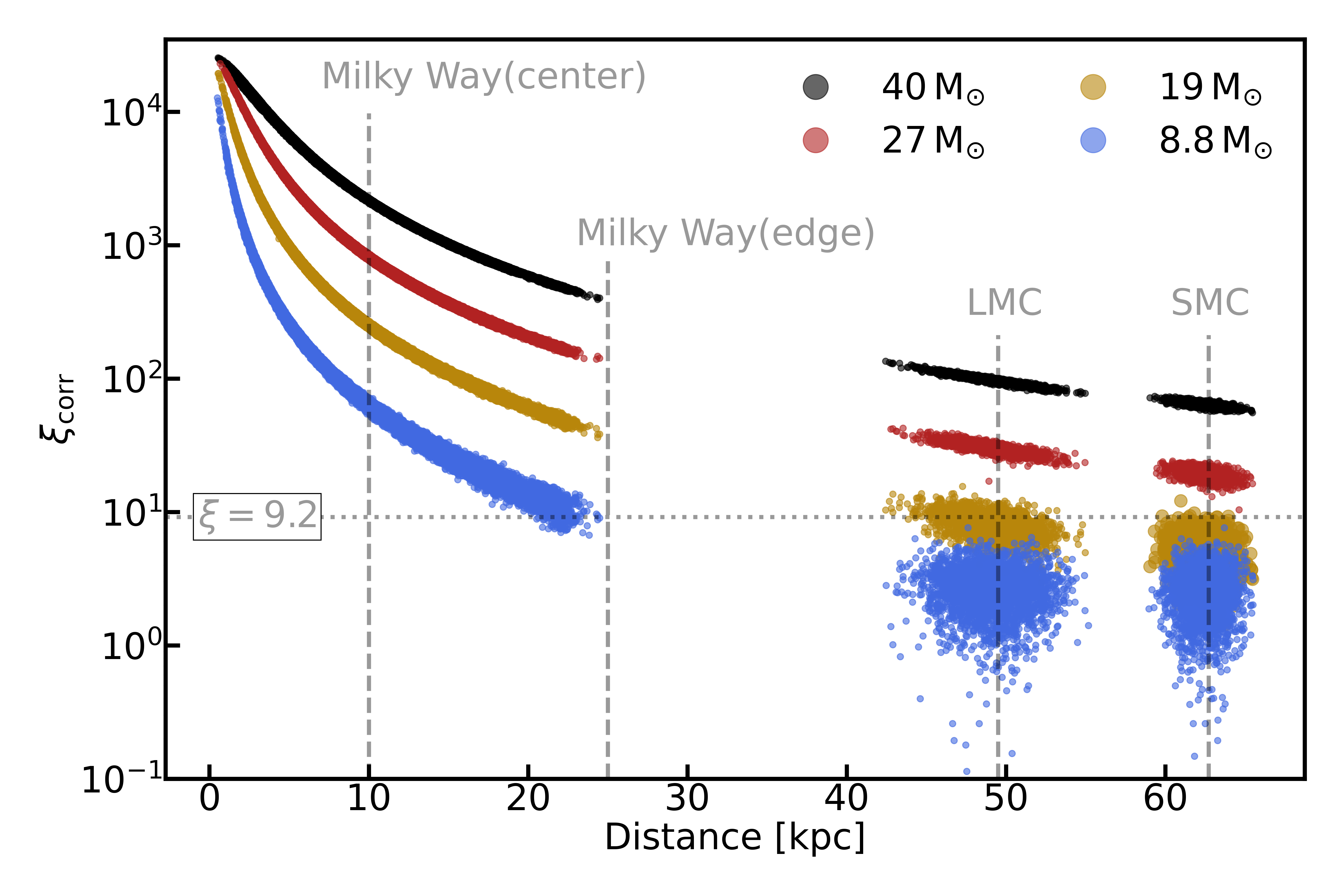}\hfill
    \includegraphics[height=6cm]{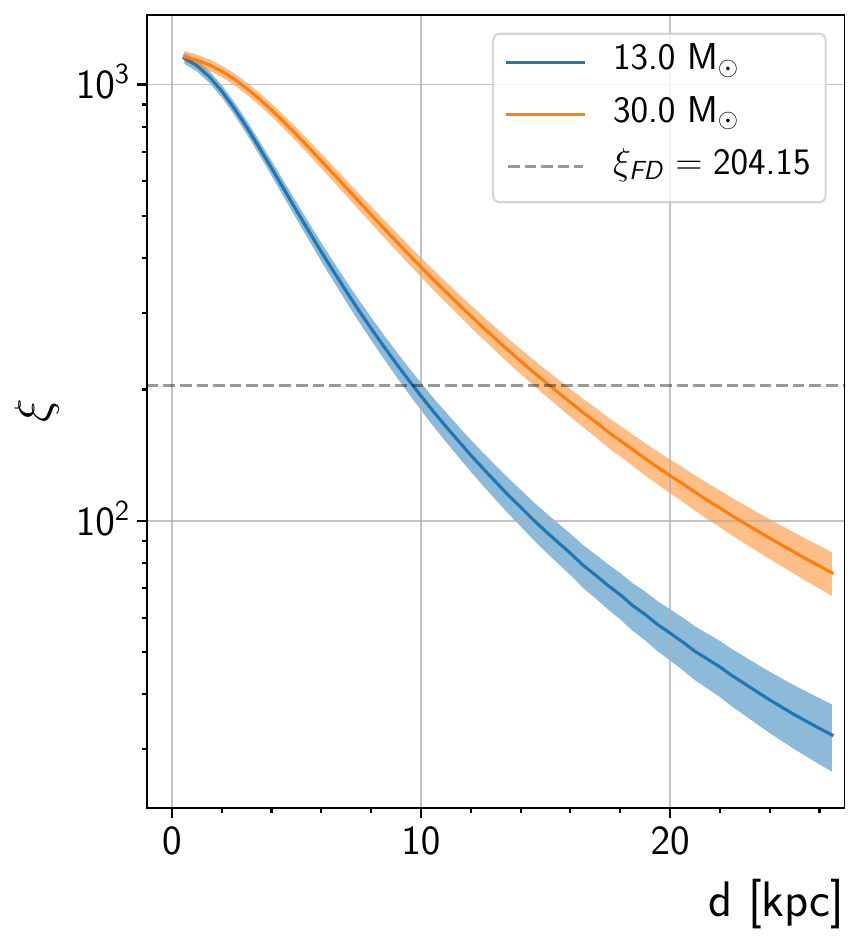}
    \caption{Test statistic $\xi$ as a function of distance $d$ (left) for a variety of CCSN progenitor models \cite{fritz_thesis, Fritz:2023}, and (right) for a 13 M$_\odot$ and 30 M$_\odot$ progenitors from \cite{Nakazato:2012qf}. A trigger with $\xi_\mathrm{FD}=204.15$ (dashed line) was created during the ``Fire Drill'' test described in Sec. \ref{sec:firedrill}.}
    \label{fig:xi_vs_dist}
\end{figure}

To search for the correlated DOM hits produced by a CCSN neutrino burst, IceCube’s Supernova Data Acquisition (SNDAQ) scans the detector hit stream in real time for significant collective deviations from the average DOM hit rate. An artificial non-paralyzing deadtime of 250 $\mu$s is applied to lower the per-DOM background rate, allowing this collective excess to be measured on top of a background of 286 Hz per DOM. 
The hit rate of each DOM is measured and reported as a continuous stream of scaler data, given by the number of hits observed in a 1.6384 ms window. The scaler streams are re-binned in software to 2 ms and a real time analysis is performed by SNDAQ. The software continuously updates the background rate for each DOM $i$ in a sliding 10-minute time window and compares it against the instantaneous rate $R_i$ within a central adjustable ``signal'' time window. A trigger is formed by summing over all DOMs and searching for statistically significant excesses above the background by maximizing the likelihood
\begin{equation}\label{eq:llh}
\mathcal{L}(\Delta \mu) = \prod^{N_\text{DOM}}_{i=1} \frac{1}{\sqrt{2\pi}\left<\sigma_i\right>} \exp\left( -\frac{(R_i - (\left<R_i\right>+\varepsilon_i \cdot \Delta\mu))^2}{2\left<\sigma_i\right>^2}\right),
\end{equation}
with $N_\text{DOM}=5,160$ and $\varepsilon_i$ giving the relative photon detection efficiency of the $i\text{th}$ DOM. For standard DOMs, $\bar{\varepsilon}_i=1.0$, and for the high quantum efficiency DOMs in the DeepCore detector, $\bar{\varepsilon}_i=1.35$. The free parameter in (\ref{eq:llh}) is $\Delta\mu$, the collective rise in the hit rate across all DOMs during the signal window. The quantity $\langle\sigma_i\rangle^2$ is the variance in the hit rate of DOM $i$ estimated using the sliding 10-minute background window. Maximizing (\ref{eq:llh}) yields
\begin{equation}\label{eq:maxllh}
    \Delta\mu = \sigma^{2}_{\Delta\mu} \sum^{N_\text{DOM}}_{i=1} \frac{\varepsilon_i (R_i - \left<R_i\right>)}{\left<\sigma_i\right>^2} \quad \text{ and}\quad \sigma^{2}_{\Delta\mu} = \left(\sum^{N_\text{DOM}}_{i=1} \frac{\varepsilon^2_i}{\left<\sigma_i\right>^2}\right)^{-1},
\end{equation}
where $\sigma^2_{\Delta\mu}$ is the estimated variance of the maximum likelihood value $\Delta\mu$. The significance of the collective rate increase is expressed in terms of a test statistic $\xi$, given by the ratio
\begin{equation}\label{eq:xi}
    \xi = \frac{\Delta\mu}{\sigma_{\Delta\mu}}.
\end{equation}
The test statistic $\xi$ is correlated with the seasonally dependent atmospheric muon rate \cite{Baum:2017rty}. This effect is corrected at the time of trigger formation, producing the so-called "corrected test statistic" $\xi^\prime$. $\xi$ and $\xi^\prime$ may be used to characterize IceCube's sensitivity to a variety of CCSN models as a function of distance, as illustrated in Fig. \ref{fig:xi_vs_dist}. 

\section{IceCube "Fire Drill" Data challenges}\label{sec:firedrill}
The detection of a Galactic CCSN will trigger an operational response that includes securing raw data, validating it, and then preparing a publication on the observation. This procedure, known as the "escalation scheme", is designed to reduce the chance of false positive identification and to verify any physics results before issuing a public announcement \cite{Griswold:2021osi}.
When a CCSN trigger forms, a series of automated steps are performed based on the values of $\xi$ and $\xi^\prime$. This includes email notifications to experts and working group leads within IceCube, data buffering requests, and issuance of an alert to SNEWS. Additional human intervention is performed for triggers with high $\xi$. When an alert is formed with $\xi>10$ or $\xi^\prime>10$, known as a ``gold'' alert, the detector operators ensure the IceCube detector is running normally, and the IceCube spokesperson, Supernova Working Group, and Executive Committee review the data and any results obtained from them. 
Given the rarity of Galactic CCSNe, it is necessary to perform regular tests of the escalation scheme to identify failure modes and edge cases where the scheme fails. A ``Fire Drill'' system was developed to test IceCube’s supernova trigger formation and automated notification distribution \cite{Griswold:2021osi}. IceCube's response to a CCSN explosion was simulated using ASTERIA \cite{ASTERIA:2019} (See Fig. \ref{fig:hits_phases}). The resulting lightcurve was used to construct a time series of hits in each of the IceCube DOMs. These were then injected into archival DOM background data captured in September 2015. The archival data were chosen from a period known to be stable when there were no known astrophysical transients in progress. The tags used by pDAQ to determine coincidence were generated for the simulated hits and were modified for the background hits, where appropriate. The result of this injection was a dataset describing IceCube's response to a CCSN at the level of individual hits.

\begin{figure}[b]
    \centering
    \includegraphics[width=0.9\textwidth]{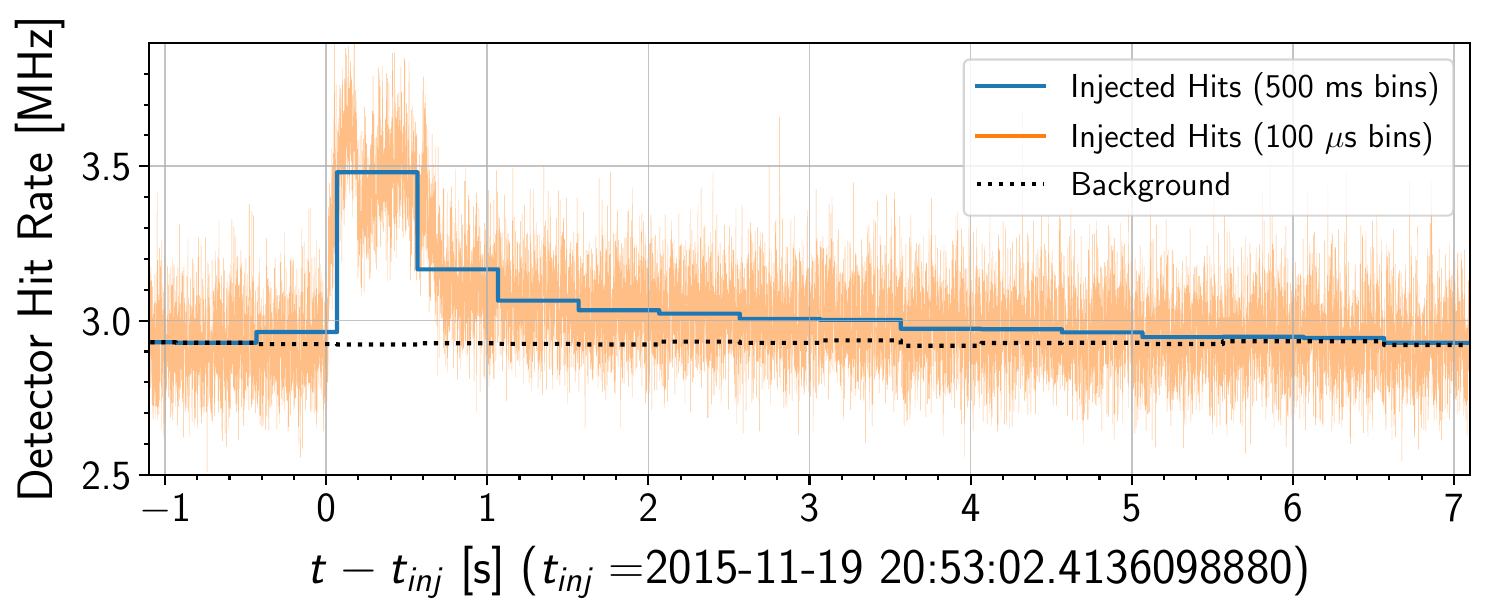}
    \caption{The injected neutrino lightcurve from the offline fire drill tests in 0.5~s bins, as if processed by SNDAQ, and in 100~$\mu$s rebinned from individual Hits, as if captured by HitSpool. This lightcurve does not include the reduction imposed by the 250 $\mu$s deadtime described in Sec. \ref{sec:sn_detection}.}
    \label{fig:fd_lightcurve}
\end{figure}

Following the injection, the modified data were staged to the IceCube South Pole Testing System (SPTS) and "replayed" as if they were being observed in the detector in real-time \cite{Aartsen:2016nxy}. They were processed by pDAQ and subsequently by SNDAQ, resulting in the formation of supernova candidate triggers. Four tests were performed during the summer of 2021 during the development of this system. The final test, described in Fig. \ref{fig:fd_lightcurve}, yielded a supernova candidate trigger with $\xi=204.15$. The expected $\xi$ corresponding to a CCSN from a 13~M$_\odot$ and a 30~M$_\odot$ progenitor is shown in Fig.~\ref{fig:xi_vs_dist}. At 10~kpc, the expectation for the 13~M$_\odot$ progenitor used during the Fire drill is $\xi=193.36\pm13.91$. 

The tests performed at SPTS were offline: any notifications or emails produced in response to triggers were suppressed. The tests were also open: the collaboration and detector operators were notified in advance. By monitoring the automated components of the escalation scheme during the offline tests, a number of low-level issues were identified and resolved \cite{Griswold:2021osi}. In the future, IceCube will perform regular blind, online fire drills: tests that issue alerts and are not announced ahead of time. This is intended to improve the readiness of the collaboration and detector operators. The system that will perform this is described in Sec. \ref{sec:future_work}.

\section{IceCube's Role in SNEWS followup}\label{sec:snews}
IceCube has been closely involved in the development of SNEWS 2.0 software. \texttt{SNEWPY} \cite{baxter2021snewpy} is integrated with IceCube’s Fast Supernova response Monte Carlo ASTERIA \cite{ASTERIA:2019}. This was used to perform the signal injection described in Sec. \ref{sec:firedrill}. IceCube has also participated in several offline fire drills using \texttt{snews\_pt} \cite{snews_pt}, including a test of SNEWS2.0's ability to triangulate the direction of a transient \cite{Kharusi:2020ovw}. 

\begin{figure}[h]
    \centering
    \includegraphics[width=0.49\textwidth]{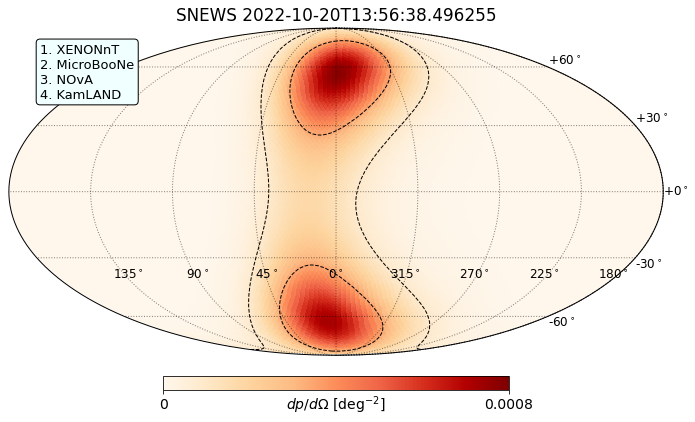}
    \includegraphics[width=0.49\textwidth]{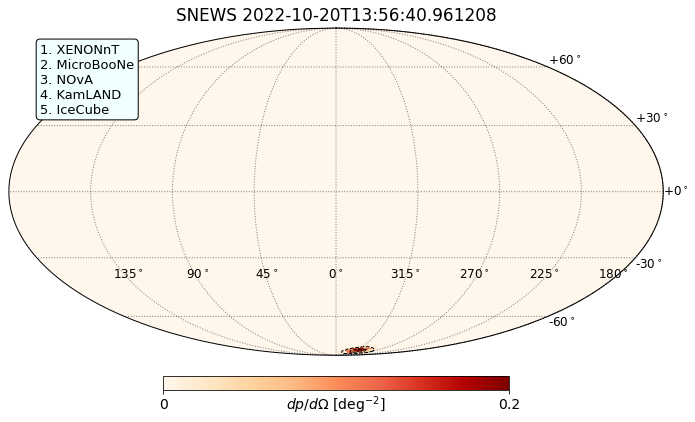}
    \caption{Credible regions in the sky, obtained from a combination of simulated observations from SNEWS 2.0-member neutrino detectors, (left) excluding IceCube, (right) including IceCube. Figures used with permission, provided at the courtesy of the SNEWS2.0 Collaboration.} 
    \label{fig:triangulation}
\end{figure}

In these tests, the arrival time of a CCSN neutrino burst at multiple detectors around Earth was simulated. These times were used to generate artificial alerts via \texttt{snews\_pt}, and the differences between them were used to triangulate the direction of the potential progenitor \cite{Coleiro_2020}. The credible regions of the sky based on the results of this test are shown in Fig. \ref{fig:triangulation}. In addition to providing a high statistic measurement of the neutrino lightcurve, IceCube's location in the Southern hemisphere and distance from other neutrino detectors makes it especially useful for the purposes of triangulation. Triangulation-based pointing is paramount to ensure robust optical follow-up.

\section{Future Work}\label{sec:future_work}
The current method of performing a signal injection fire drill at IceCube can only be performed on archival data.  
Re-processing archival data requires a temporary interruption of the regular data flow, which would obstruct other experimental efforts in the context of an online test. An improved signal injection system that operates in parallel with IceCube's regular data flow is under development. This system will accompany a major upgrade to IceCube’s SNDAQ and is expected to deploy in late 2023. The method of performing a fire drill using this new system will be as follows: (1) schedule a fire drill hours to days in advance; (2) using a pre-simulated CCSN neutrino lightcurve, generate a stream of scaler data in parallel with SNDAQ’s real data stream; (3) at the scheduled time, combine the real and simulated scaler data streams via summation over matching time bins; (4) form SNDAQ triggers and issue alerts as appropriate; and (5) separate the real and simulated hits so as to preserve the real data's integrity. As the timing of true SNDAQ triggers cannot be known in advance, a scheduled fire drill can be aborted based on the status of the detector and SNDAQ at the time of the test's start. This system may be used for both online and offline fire drills and is highly configurable, enabling tests with or without SNEWS alerts, blind or open data challenges, and with a variety of CCSN models provided by \texttt{SNEWPY}.

As development on \texttt{snews\_pt} continues, the scope of SNEWS 2.0 fire drills will also expand. One thrust of SNEWS 2.0’s mission is the involvement of amateur astronomers. Future fire drills with \texttt{snews\_pt} will eventually involve the issuance of artificial CCSN alerts to amateur astronomer groups, such as the American Association of Variable Star Observers, and subscribers to SNEWS mailing lists.

\bibliographystyle{ICRC}
\bibliography{references}

\clearpage

\section*{Full Author List: IceCube Collaboration}

\scriptsize
\noindent
R. Abbasi$^{17}$,
M. Ackermann$^{63}$,
J. Adams$^{18}$,
S. K. Agarwalla$^{40,\: 64}$,
J. A. Aguilar$^{12}$,
M. Ahlers$^{22}$,
J.M. Alameddine$^{23}$,
N. M. Amin$^{44}$,
K. Andeen$^{42}$,
G. Anton$^{26}$
% ...
\\
\\
% ...
$^{12}$ Universit{\'e} Libre de Bruxelles, Science Faculty CP230, B-1050 Brussels, Belgium \\
% ...
$^{17}$ Department of Physics, Loyola University Chicago, Chicago, IL 60660, USA \\
$^{18}$ Dept. of Physics and Astronomy, University of Canterbury, Private Bag 4800, Christchurch, New Zealand \\
% ...
$^{22}$ Niels Bohr Institute, University of Copenhagen, DK-2100 Copenhagen, Denmark \\
$^{23}$ Dept. of Physics, TU Dortmund University, D-44221 Dortmund, Germany \\
% ...
$^{26}$ Erlangen Centre for Astroparticle Physics, Friedrich-Alexander-Universit{\"a}t Erlangen-N{\"u}rnberg, D-91058 Erlangen, Germany \\
% ...
$^{40}$ Dept. of Physics and Wisconsin IceCube Particle Astrophysics Center, University of Wisconsin{\textendash}Madison, Madison, WI 53706, USA \\
% ...
$^{42}$ Department of Physics, Marquette University, Milwaukee, WI, 53201, USA \\
% ...
$^{44}$ Bartol Research Institute and Dept. of Physics and Astronomy, University of Delaware, Newark, DE 19716, USA \\
% ...
$^{63}$ Deutsches Elektronen-Synchrotron DESY, Platanenallee 6, 15738 Zeuthen, Germany  \\
$^{64}$ Institute of Physics, Sachivalaya Marg, Sainik School Post, Bhubaneswar 751005, India \\
% ...

\subsection*{Acknowledgements}

\noindent
The authors gratefully acknowledge the support from the following agencies and institutions:
USA {\textendash} U.S. National Science Foundation-Office of Polar Programs,
U.S. National Science Foundation-Physics Division,
U.S. National Science Foundation-EPSCoR,
Wisconsin Alumni Research Foundation,
Center for High Throughput Computing (CHTC) at the University of Wisconsin{\textendash}Madison,
Open Science Grid (OSG),
Advanced Cyberinfrastructure Coordination Ecosystem: Services {\&} Support (ACCESS),
Frontera computing project at the Texas Advanced Computing Center,
U.S. Department of Energy-National Energy Research Scientific Computing Center,
Particle astrophysics research computing center at the University of Maryland,
Institute for Cyber-Enabled Research at Michigan State University,
and Astroparticle physics computational facility at Marquette University;
Belgium {\textendash} Funds for Scientific Research (FRS-FNRS and FWO),
FWO Odysseus and Big Science programmes,
and Belgian Federal Science Policy Office (Belspo);
Germany {\textendash} Bundesministerium f{\"u}r Bildung und Forschung (BMBF),
Deutsche Forschungsgemeinschaft (DFG),
Helmholtz Alliance for Astroparticle Physics (HAP),
Initiative and Networking Fund of the Helmholtz Association,
Deutsches Elektronen Synchrotron (DESY),
and High Performance Computing cluster of the RWTH Aachen;
Sweden {\textendash} Swedish Research Council,
Swedish Polar Research Secretariat,
Swedish National Infrastructure for Computing (SNIC),
and Knut and Alice Wallenberg Foundation;
European Union {\textendash} EGI Advanced Computing for research;
Australia {\textendash} Australian Research Council;
Canada {\textendash} Natural Sciences and Engineering Research Council of Canada,
Calcul Qu{\'e}bec, Compute Ontario, Canada Foundation for Innovation, WestGrid, and Compute Canada;
Denmark {\textendash} Villum Fonden, Carlsberg Foundation, and European Commission;
New Zealand {\textendash} Marsden Fund;
Japan {\textendash} Japan Society for Promotion of Science (JSPS)
and Institute for Global Prominent Research (IGPR) of Chiba University;
Korea {\textendash} National Research Foundation of Korea (NRF);
Switzerland {\textendash} Swiss National Science Foundation (SNSF);
United Kingdom {\textendash} Department of Physics, University of Oxford.

\end{document}